\documentclass[aps,prl,showpacs,twocolumn]{revtex4-2}
\usepackage{bm}
\usepackage{amsmath}
\usepackage{amsfonts}
\usepackage{amssymb}
\usepackage{times}
\usepackage{siunitx}
\usepackage{graphicx}
\usepackage{dcolumn}
\usepackage{bm}
\usepackage{dcolumn}
\newcolumntype{d}[1]{D{.}{.}{#1}}

\newcommand*{\centt}[1]{\multicolumn{1}{c}{#1}}

\allowdisplaybreaks

\usepackage{xcolor}

\newcommand\mathtight
{
\thinmuskip=1mu   
\medmuskip=2mu    
\thickmuskip=3mu  
}

\begin{document}
\preprint{Version 1.0}

\title{QED effects on the nuclear magnetic shielding of $^3$He}

\author{Dominik Wehrli}
\affiliation{Laboratorium f\"ur Physikalische Chemie, ETH-Z\"urich, 8093 Z\"urich, Switzerland}

\author{Anna Spyszkiewicz-Kaczmarek}
\affiliation{Faculty of Chemistry, Adam Mickiewicz University, Uniwersytetu Pozna{\'n}skiego 8, 61-614 Pozna{\'n}, Poland}

\author{Mariusz Puchalski}
\affiliation{Faculty of Chemistry, Adam Mickiewicz University, Uniwersytetu Pozna{\'n}skiego 8, 61-614 Pozna{\'n}, Poland}

\author{Krzysztof Pachucki}

\affiliation{Faculty of Physics, University of Warsaw,
             Pasteura 5, 02-093 Warsaw, Poland}

\date{\today}

\begin{abstract}
The leading quantum electrodynamic corrections to the nuclear magnetic shielding in one- and two-electron atomic systems 
are obtained in a complete form, and the shielding constants of $^1$H, $^3$He$^+$, and $^3$He are calculated to be $17.735\,436(3) \cdot 10^{-6}$, $35.507\,434(9)\cdot 10^{-6}$, and $59.967\,029(23)\cdot 10^{-6}$, respectively.
These results are orders of magnitude more accurate than previous ones, and, 
with the ongoing measurement of the nuclear magnetic moment of $^3$He$^+$ and planned $^3$He$^{2+}$,
they open the window for high-precision absolute magnetometry using $^3$He NMR probes. 
The presented theoretical approach is applicable to all other light atomic and molecular systems, 
which facilitates the improved determination of magnetic moments of any light nuclei. 
\end{abstract}

\maketitle

{\sl Introduction} -- The chemical inertness of $^3$He atomic gas and the absence of an electric quadrupole moment
yield an exceptional isolation of nuclear spins that form a macroscopic quantum state achieving long coherence times. Because it can be hyper-polarized using laser optical pumping techniques, resulting in a very stable and sensitive probe 
to the magnetic field, the nuclear magnetic resonance (NMR) magnetometry founded on $^3$He is of great interest 
for different areas of physics and applied sciences \cite{Gentile:2017}. In particular, $^3$He gas cells have recently been used 
in several fundamental physics experiments for their capability of determining absolute field values, in the calibration of magnetic sensors, e.g., in muon $g-2$ measurements \cite{Abi2021}, in searching for a permanent electric dipole moment \cite{Sachdeva:2019}, 
and in the development of cryogenic NMR techniques for new experiments on the electron and positron magnetic moments \cite{Gabrielse:2019, Fan:2019}. For this reason, hyper-polarized $^3$He NMR probes have been proposed as a new standard for absolute magnetometry \cite{Nikiel:2014,farooq:2020, Fan:2019}, and consequently, a high-precision value of the helion nuclear magnetic moment $\mu_h$ would be indispensable.

For the determination of $\mu_h$, the NMR measurements of $^3$He performed with respect to H$_2$ and H$_2$O \cite{neronov14,flowers1993,aruev2012,jackowski2010,garbacz2012}
have been used to obtain the current CODATA value \cite{tiesinga21} with a relative accuracy of $1.2\cdot 10^{-8}$. 
This approach, based on a comparison to the proton magnetic moment $\mu_p$, is limited by the little-known 
magnetic shielding effects caused by the surrounding particles (electrons, nuclei). To obtain accurate shielding factors, we can consider using theoretical methods. Nonetheless, this possibility only applies to small atomic and molecular systems, for which we can use an approach based on nonrelativistic implementation of quantum electrodynamics (NRQED). Such calculations in H$_2$ with an accuracy similar to the magnetic shielding of $^3$He performed in this work can lead to $\mu_h$ with relative accuracy as high as of that of $\mu_p$, i.e., 2.9 $\cdot 10^{-10}$ \cite{schneider19}. 

In another approach, the $^3$He magnetic moment is related to that of the electron. This is achieved by measuring the ratio of the magnetic moment of the $2^3S$ metastable state of $^4$He to the ground state of $^3$He \cite{shifrin97}, and, independently, by a very recent measurement of the Breit-Rabi splitting of $^3$He$^+$ in a Penning trap \cite{mooser18, private}. This latter experiment may provide an improved result for $\mu_h$ in comparison to the current CODATA value. 

Finally, an effort is underway \cite{mooser18,schneider19} to directly measure $\mu_h$ in a cryogenic Penning trap using techniques similar to those applied for the proton \cite{mooser14,schneider17} and the antiproton \cite{smorra17}. The high-precision results of the magnetic shielding constant obtained in this work provide the shielded magnetic moment of $^3$He$^+$ and $^3$He at the accuracy of the measured value of $\mu_h$, which is a prerequisite for the realization of $^3$He NMR probes as a new standard for absolute magnetometry.


{\sl Magnetic shielding} -- Let us recall the definition of the nuclear diamagnetic shielding of a single atom \cite{Lamb:1941,ramsey50}. The coupling of the nuclear magnetic moment $\vec\mu$ with the homogeneous magnetic field $\vec B$ modified by the presence of atomic electrons can be parametrized in terms of the nuclear magnetic shielding constant $\sigma$ according to
\begin{equation}
\delta H = -\vec\mu\cdot\vec B\,(1-\sigma). \label{eq:shielding_definition}
\end{equation}
In order to calculate accurately the parameter $\sigma$ we employ the so-called nonrelativistic QED \cite{caswell86}, and assume the expansion of $\sigma$ as a double power series in the fine structure constant $\alpha$ and the electron-nucleus mass ratio $m/m_N$,
\begingroup
\mathtight
\begin{align}
\sigma &= \sigma^{(2)} + \sigma^{(4)} + \sigma^{(5)} + \sigma^{(6)}  +\sigma^{(2,1)} + \sigma^{(2,2)} + \sigma^{(4,1)} + \ldots \label{eq:sigma}
\end{align}
\endgroup
In this equation, $\sigma^{(n)}\propto\alpha^n$ coefficients are the nonrelativistic shielding,
the relativistic, the leading QED, and the higher-order QED corrections in the infinite nuclear mass approximation, respectively. The terms $\sigma^{(n,k)}\propto\alpha^n\,(m/m_N)^k$ are corrections due to the finite nuclear mass. The main advantage of this expansion is the possibility to derive exact formulas for expansion coefficients in terms of some matrix element with the nonrelativistic wave function, which can be accurately calculated.

In the case of one- and two-electron atomic systems all lower-order contributions in Eq.~\eqref{eq:sigma} are known in the literature. Namely, the formula for the leading term $\sigma^{(2)}$ valid for atoms 
\begin{align}
\sigma^{(2)} = \frac{\alpha}{3\,m}\,\Bigl\langle\sum_a \frac{1}{r_a}\Bigr\rangle \label{03}
\end{align}
was first introduced by Lamb~\cite{Lamb:1941} and later generalized for molecules by Ramsey~\cite{ramsey50}. The correction $\sigma^{(4)}$ can be found in Refs.~\cite{ivanov09,rudzinski09} and references therein, while  $\sigma^{(2,1)}$ and $\sigma^{(2,2)}$ were derived in Ref.~\cite {pachucki08}. In contrast, QED corrections until now have only been partially investigated and therefore are presently the bottleneck of theoretical predictions. 
There have been attempts \cite{gimenez16} to include them within  the formalism based on the Dirac-Coulomb (DC) Hamiltonian \cite{Kutzelnigg:2012},
but there is currently no adequate formulation of the QED theory for many electron systems. 
While, for hydrogenic ions, Yerokhin {\em et al.} \cite{yerokhin11,yerokhin12} performed a nonperturbative numerical evaluation of one-loop QED contributions and observed a slow numerical convergence for the small nuclear charge $Z$. Therefore, these results were supplemented by direct NRQED evaluation of the leading QED correction $\sim\alpha^5$. However, some effects due to the magnetic moment anomaly were omitted there, which was the reason for small discrepancies with the nonperturbative results for the medium-Z hydrogen-like ions. For helium, only the leading QED logarithmic correction $\sim \alpha^5\,\ln\alpha$ was obtained by Rudzi\'nski {\em et al.} in Ref.~\cite{rudzinski09}. In this Letter, we present the complete formulas for $\sigma^{(5)}$, valid for one- and two-electron atomic systems, which can be easily generalized to any light few-electron system. They are supplemented by 
the numerical calculations performed for $^1$H, $^3$He$^+$ and $^3$He, using explicitly correlated exponential functions for the latter case,
what ensures high precision of numerical results. 
We will use $\hbar=c=\varepsilon_0=1$ throughout the paper and the CODATA 2018 values of physical constants \cite{tiesinga21}.

{\sl NRQED approach} -- In the NRQED formalism we can include QED effects, coming from large photon momenta, 
    in the generalized Breit-Pauli Hamiltonian (see e.g. \cite{pachucki04}). 
For the case of a two-electron atomic system coupled to a magnetic field $\vec{B}$, this effective Hamiltonian is given by
\begin{equation}
H_\mathrm{BP}=\sum_{a=1}^2 H_a + H_{12},\label{breit_gen} 
\end{equation}
\begin{widetext}
\begingroup
\mathtight
\begin{align}
H_a &= \frac{\vec{\pi}_a^2}{2\,m} -\frac{\vec{\pi}_a^4}{8\,m^3} -\frac{Z\,\alpha}{r_a} -\frac{e\,(1+\kappa)}{2\,m}\,\vec{\sigma}_a\cdot\vec{B}_a +\frac{e}{8\,m^3}\,\lbrace\vec{\pi}_a^2,\vec{\sigma}_a\cdot\vec{B}_a\rbrace -\frac{e^2}{2}\bigg(\frac{1}{4\,m^3}+\alpha_M\bigg)\vec{B}_a^2
+\frac{e\,\kappa}{8\,m^3}\,\lbrace\vec{\pi}_a\cdot\vec{B}_a,\vec{\sigma}_a\cdot\vec{\pi}_a\rbrace
\nonumber\\
{}& \hspace{-0.5cm} +\frac{Z\,\alpha\,(1+2\,\kappa)}{4\,m^2}\,\vec{\sigma}_a\cdot\frac{\vec{r}_a\times\vec{\pi}_a}{r_a^3}
+\frac{e}{6\,m}\,\bigg(r^2_E+r^2_{\rm vp}-\frac{3\,\kappa}{4\,m^2}\bigg)(\nabla_a\times \vec{B}_a)\cdot\vec{\pi}_a +\frac{2\,\pi\,Z\,\alpha}{3}\bigg(\frac{3}{4\,m^2}+r^2_E+r^2_{\rm vp}\bigg)\,\delta(\vec{r_a}), \\
H_{12}&=\frac{\alpha}{r}-\frac{4\,\pi\,\alpha}{3}\left(\frac{3}{4\,m^2}+r_E^2 + \frac{1}{2}\,r^2_{\rm vp}
+\frac{\,(1+\kappa)^2}{2\,m^2}\,\vec{\sigma}_1\cdot\vec{\sigma}_2\right)\delta(\vec{r})
+ \frac{\alpha}{4\,m^2\,r^3}\Big[2\,(1+\kappa)\big(\vec{\sigma}_1\cdot\vec{r}\times\vec{\pi}_2-\vec{\sigma}_2\cdot\vec{r}\times\vec{\pi}_1\big) \nonumber\\ 
&\ \hspace{-0.5cm} +(1+2\,\kappa)\big(\vec{\sigma}_2\cdot\vec{r}\times\vec{\pi}_2-\vec{\sigma}_1\cdot\vec{r}\times\vec{\pi}_1\big)\Big]
- \frac{\alpha}{2\,m^2}\,\pi_1^i\left(\frac{\delta^{ij}}{r}+\frac{r^i\,r^j}{r^3}\right)\pi_2^j
-\frac{3\,\alpha\,(1+\kappa)^2}{4\,m^2}\,\frac{\sigma_1^i\,\sigma_2^j}{r^3} \biggl(\frac{r_1^i\,r_2^j}{r^2}-\frac{\delta^{ij}}{3}\biggr) 
\label{eq:breithamiltonian}
\end{align}
\endgroup
\end{widetext}
where $\vec{r}=\vec{r}_1-\vec{r}_2$, $\vec \pi = \vec p - e\,\vec A$ is the generalized momentum in the external field, $\kappa=\alpha/(2\,\pi)$ is the magnetic moment anomaly, and
\begin{align}
r_E^2&=\frac{3\,\kappa}{2\,m^2}+6\,F'_1(0) = 
\frac{2\,\alpha}{\pi\,m^2}\left(\ln\frac{m}{2\,\epsilon}+\frac{5}{6}\right),\label{eq:re2}\\
r^2_{\rm vp} &= -\frac{2\,\alpha}{5\,\pi\,m^2},\label{eq:rvp}\\
\alpha_M&=\frac{4\,\alpha}{3\,\pi\, m^3} \left(-\ln\frac{m}{2\,\epsilon}+\frac{13}{24}\right).\label{eq:alpham}
\end{align}
The parameters $r_E$ and $\alpha_M$ are interpreted as the charge radius and the magnetic polarizability of an electron, 
respectively. They depend on $\epsilon$, i.e. the photon momentum cut-off being used as a regulator \cite{itzykson05}, and this dependence cancels out in the complete expression for any physical quantity. The formula for $r^2_E$ is derived from the known radiative correction to electromagnetic form factors $F_1$ and $F_2$ \cite{itzykson05}, and $r^2_{\rm vp}$ incorporates the corrections due to the vacuum polarization. The formula for $\alpha_M$ has for the first time been presented in Ref. \cite{yerokhin12} and can be obtained in a similar way as for the electric polarizability, denoted by $\chi$, in Ref. \cite{jentschura05}. Without QED ($r^2_E=r^2_{\rm vp}=\alpha_M=\kappa=0$) $H_\mathrm{BP}$ is the standard Breit-Pauli Hamiltonian \cite{bethe77}. 

For the derivation of the magnetic shielding constant $\sigma$, we consider Eq.~\eqref{breit_gen} for an atomic system in 
a magnetic field corresponding to the sum of the vector potential $\vec A_I$ due to the magnetic moment $\vec\mu$ of the nucleus , and $\vec A_E$ due to the homogeneous external magnetic field $\vec{B}$, namely 
\begin{equation}
\vec A_E = \frac{1}{2}\,\vec B\times\vec r,\quad {\rm and} \quad \vec A_I = \frac{1}{4\,\pi}\,\vec\mu\times\frac{\vec r}{r^3}.
\end{equation}
Following Ramsey's theory of the magnetic shielding \cite{ramsey50,helgaker99}, we split the Hamiltonian $H_\mathrm{BP}$ as
\begin{align}
H_\mathrm{BP}&= H_0 + \delta H_{\vec{A}_E=\vec{A}_I=0}+ \delta H_{\vec{A}_I,\vec{A}_E=0}+ \delta H_{\vec{A}_E,\vec{A}_I=0}
\nonumber \\ 
& +\delta H_{\vec{A}_E,\vec{A}_I}+O(\vec{A}^2_{I,E}),
\end{align}
where $\delta H$ is treated as a perturbation to the nonrelativistic Hamiltonian $H_0$,
$\delta H_{\vec{A}_E=\vec{A}_I=0}$ is independent of the magnetic fields,
$\delta H_{\vec{A}_I,\vec{A}_E=0}$ is linear in $\vec{A}_I$, $\delta H_{\vec{A}_E,\vec{A}_I=0}$
is linear in $\vec{A}_E$, and $\delta H_{\vec{A}_E,\vec{A}_I}$ is bilinear in both fields.
Because we are only interested in energy corrections that are proportional to $\vec{\mu}\cdot\vec{B}$, we write
\begingroup
\mathtight
\begin{align}
\delta E &= \langle\delta H_{\vec{A}_E,\vec{A}_I}\rangle + 2\left\langle\delta H_{\vec{A}_E,\vec{A}_I}\,\frac{1}{(E_0-H_0)'}\,\delta H_{\vec{A}_E=\vec{A}_I=0}\right\rangle\nonumber \\ 
&+ 2\left\langle\delta H_{\vec{A}_I,\vec{A}_E=0}\,\frac{1}{(E_0-H_0)'}\,\delta H_{\vec{A}_E,\vec{A}_I=0}\right\rangle+\ldots \,, 
\label{eq:ecorrectiongeneral}
\end{align}
\endgroup
where $1/(E_0-H_0)'$ is the reduced Green's function, and the ellipses denote terms that are not proportional to $\vec{\mu}\cdot\vec{B}$ and will be discarded. The expectation values are taken with respect to the ground electronic state of $H_0$, which is an $S$ state in the case of hydrogen- and helium-like systems. The spherical symmetry then implies the relation $\mu^iB^j= \delta^{ij}/3\,\vec{\mu}\cdot\vec{B}$ allowing for a simple factoring of $\vec{\mu}\cdot\vec{B}$ from many terms appearing in Eq.\,\eqref{eq:ecorrectiongeneral}, and $\sigma$ is obtained
through the relation $\delta E = \vec{\mu}\cdot\vec{B}\,\sigma$.


{\sl Leading QED correcion} -- We derive the correction $\sigma^{(5)}$ for a helium-like system, bearing in mind the corresponding derivation of the Lamb shift \cite{pachucki98}. Namely, $\sigma^{(5)}$ is given by the sum of the high $\sigma_B$ and the low 
$\sigma_A$  energy parts
\begin{equation}
\sigma^{(5)} = \sigma_{B0} + \sigma_{B1} +  \sigma_{B2} +  \sigma_{B3} +  \sigma_{A1} +  \sigma_{A2}. \label{eq:sigma5}
\end{equation}
The high energy part is obtained as follows.
$\sigma_{B0}$ corresponds to the helium Lamb shift \cite{pachucki98} with the wave function corrected by the effect of magnetic fields 
in Eq. (\ref{03})
\begingroup
\mathtight
\begin{align}
\sigma_{B0} &= \frac{2\,\alpha^3}{3\,m^3}\,\biggl\langle{\mkern-3mu}\biggl(\frac{1}{r_1}+\frac{1}{r_2}\biggr)\frac{1}{(E_0-H_0)'}\biggl\{
{\mkern-3mu}\left(\frac{19}{30}+\ln(\alpha^{-2})\right)\nonumber\\
& \times \frac{4\,Z}{3}\,(\delta(\vec{r_1})+\delta(\vec{r_2})) 
+ \left(\frac{164}{15}+\frac{14}{3}\,\ln\alpha\right)\delta(\vec{r})\nonumber\\  
&  -\frac{7\,\alpha^3\,m^3}{6\,\pi}\,P\left(\frac{1}{(m\,\alpha\, r)^3}\right){\mkern-3mu}\biggr\}{\mkern-3mu}\biggr\rangle.
\end{align}
\endgroup
where $P(1/r^3)$ is the Araki-Sucher term \cite{pachucki98}. The next parts are directly obtained from the NRQED Hamiltonian in Eq.~\eqref{breit_gen}, and are the following
\begingroup
\mathtight
\begin{align}
\sigma_{B1} &= \frac{\alpha^2}{m^3}\left(\frac{20}{9}\,\ln(\alpha^{-2})-\frac{1361}{540}\right)\left\langle\delta(\vec{r}_1) + \delta(\vec{r}_2)\right\rangle, \label{eq:sigmaB1} \\  
\sigma_{B2} &= -\frac{\alpha^2}{3\,m^3} \bigg\langle{\mkern-5mu}(\delta(\vec{r}_1)-\delta(\vec{r}_2))\frac{1}{(E_0-H_0)} \label{eq:sigmaB2}\\
& \times \bigg(\frac{4\,\vec{p}_1^{\;2}}{3\,m}-\frac{4\,\vec{p}_2^{\;2}}{3\,m} -\frac{Z\,\alpha}{r_1}+\frac{Z\,\alpha}{r_2}-\frac{\alpha}{3\,r^3}\,\vec{r}\cdot(\vec{r}_1+\vec{r}_2)\bigg){\mkern-5mu}\bigg\rangle, \nonumber \\ 
\sigma_{B3} &=-\frac{3\,\alpha^2}{16\,\pi\, m^3}\bigg\langle{\mkern-5mu}\bigg(\frac{r_1^i\,r_1^j}{r_1^5}-\frac{r_2^i\,r_2^j}{r_2^5}\bigg)^{(2)}{\mkern-20mu}\frac{1}{(E_0-H_0)} \bigg( Z\,\alpha\,\frac{r_1^i\,r_1^j}{r_1^3} \nonumber\\
& \hspace{-0.3cm} -Z\,\alpha\,\frac{r_2^i\,r_2^j}{r_2^3} +\alpha\,\frac{r^i\,(r_1^j+r_2^j)}{3\,r^3}+\frac{2}{3\,m}\,(p_1^i\,p_1^j-p_2^i\,p_2^j)\bigg){\mkern-5mu}\bigg\rangle.  \label{eq:sigmaB3} 
\end{align}
\endgroup
where the second rank tensor is defined by $(p^iq^j)^{(2)}\equiv(p^iq^j + p^jq^i)/2-\delta^{ij}/3\,\vec{p}\cdot\vec{q}$. 

Analogously to the calculation of the Lamb shift \cite{pachucki98}, we define the low-energy energy contribution as
\begin{equation}
\mathtight
\delta E_A = -\frac{2\,\alpha}{3\,\pi}\left\langle{\mkern-5mu}(\vec \pi_1+\vec \pi_2)\,(H-E)\ln\frac{2\,(H-E)}{m\,\alpha^2}\,(\vec \pi_1+\vec \pi_2){\mkern-5mu}\right\rangle_{{\mkern-6mu}B}{\mkern-6mu}, \label{eq:EA}
\end{equation}
where $\langle\ldots\rangle_{B}$ denotes the expectation value with respect to the ground state with energy $E$ of the Hamiltonian in the  presence of the magnetic field
\begin{align}
H&=\frac{\vec{\pi}^2_1}{2\,m}+\frac{\vec{\pi}^2_2}{2\,m}-\frac{Z\,\alpha}{r_1}-\frac{Z\,\alpha}{r_2}+\frac{\alpha}{r} \nonumber \\
&= H_0+\frac{1}{3\,m}\,\vec{\mu}\cdot\vec{B}\,U  -\frac{e}{2\,m}\,\vec{L}\cdot\vec{B} -\frac{e}{4\,\pi\, m}\vec\mu\cdot \vec U, \label{eq:Hexp}
\end{align}
where $L^i =L_1^i+L_2^i$, $U = \alpha/r_1 + \alpha/r_2$, and $U^i =L_1^i/r_1^3+L_2^i/r_2^3$.
Equation\,\eqref{eq:EA} is only a formal expression for $\delta E_{A}$, and it needs to be expanded in the magnetic field. For this we rewrite $\delta E_{A}$ in the form
\begin{equation}
\mathtight
\delta E_{A} = -\frac{2\,\alpha}{3\,\pi}\left\langle{\mkern-5mu}(\vec r_1+\vec r_2)\,(H-E)^3\ln\frac{2\,(H-E)}{m\,\alpha^2}\,(\vec r_1+\vec r_2){\mkern-5mu}\right\rangle_{{\mkern-6mu}B}{\mkern-6mu},
\end{equation}
%
and divide $\delta E_{A}$ into two parts coming from different perturbations to $H_0$ in Eq.~\eqref{eq:Hexp}. The part due to $1/(3\,m)\,\vec{\mu}\cdot\vec{B}\, U$ leads to the first correction 
\begingroup
\mathtight
\begin{align}
\sigma_{A1} &= -\frac{2\,\alpha}{9\,\pi}\,\delta_{U}\bigg\langle{\mkern-5mu}(\vec r_1+\vec r_2)\,(H_0-E_0)^3\ln\frac{2\,(H_0-E_0)}{m\,\alpha^2}\nonumber \\
& \times (\vec r_1+\vec r_2){\mkern-5mu}\bigg\rangle = \frac{2\,\alpha}{9\,\pi\, m^2}\,{\cal D}_{A1}\,\ln k_{A1}, \label{eq:sigmaA1} \\
{\cal D}_{A1} &\equiv \ -4\, \pi\,Z \bigg\langle U\, \frac{1}{(E_0-H_0)'}\, (\delta(\vec{r_1})+\delta(\vec{r_2})) \bigg\rangle  \nonumber  \\
& + 2\, \pi\,\alpha\langle \delta(\vec{r_1})+\delta(\vec{r_2}) \rangle
\end{align}
\endgroup
The second correction is due to perturbation from the two other terms in Eq.\,\eqref{eq:Hexp},
\begin{align}
\sigma_{A2} &= -\frac{\alpha^2}{9\,\pi\,m^2}\delta_{L^i,U^i}\bigg\langle{\mkern-5mu}(\vec r_1+\vec r_2)(H_0-E_0)^3\ln\frac{2\,(H_0-E_0)}{m\,\alpha^2} \nonumber \\
& \times (\vec r_1+\vec r_2){\mkern-5mu}\bigg\rangle = -\frac{\alpha^2}{9\,\pi\,m^2}\,(1+3\,\ln k_{A2})\, {\cal D}_{A2}, \label{eq:sigmaA2}\\
{\cal D}_{A2} &\equiv 8\,\pi\langle \delta(\vec{r_1})+\delta(\vec{r_2}) \rangle
\end{align}
Both $\sigma_{A1}$ and $\sigma_{A2}$ are expressed with the Bethe-type logarithms $\ln k_{A1}$ and $\ln k_{A2}$, respectively. Their numerical calculation is not straightforward because it involves the logarithm of the Hamiltonian (see e.g. \cite{puchalski:2013}). Like the standard Bethe logarithm $\ln k_0 $, all these elements have the striking property that they are only slightly dependent on the number of electrons, as can be seen from the numerical results in the Table~\ref{tab:numvalueshe}. 


{\sl Hydrogenic formula} -- The expression $\sigma^{(5)}$ for a one-electron atomic system is obtained by direct reduction of one of the electrons in Eq.~\eqref{eq:sigma5} and related expressions. As a result, the total magnetic shielding for hydrogen-like ions, including contributions up to order $\alpha^5$, has a compact structure, namely
\begingroup
\mathtight
\begin{align}
\sigma &= \frac{1}{3}\,\alpha\,(Z\,\alpha)+\frac{97}{108}\,\alpha\,(Z\,\alpha)^3 + \frac{8}{9\,\pi}\,\alpha^2\,(Z\,\alpha)^3\,\biggl[\ln\left[(Z\,\alpha)^{-2}\right] \nonumber \\ 
& +2\ln k_0(H)-3\ln k_3 - \frac{221}{64}+\frac{3}{5}\biggr] + \frac{Z\,\alpha^2}{3}\,\biggl( \frac{1-g_N}{g_N}  - 1\biggr)\frac{m}{m_N}\nonumber\\
&+\frac{Z\,\alpha^2}{3}\left(1+Z\,\frac{2+3\,g_N}{2\,g_N}-2\,\frac{1-g_N}{g_N}\right)\frac{m^2}{m_N^2}. \label{eq:sigmaH}
\end{align}
\endgroup
where the nuclear $g$-factor is defined as 
\begin{equation}
g_N=\frac{m_N}{Z\,m_p}\,\frac{\mu}{\mu_N}\,\frac{1}{I}.
\label{eq:nuc_gfactor}
\end{equation}
In Eq.\,\eqref{eq:nuc_gfactor}, $m_p$ is the proton mass, $\mu_N$ is the nuclear magneton, and $\mu$ and $I$ are the magnetic moment 
and the spin of the considered nucleus, respectively. Moreover, the $\ln k_{A1}$ and $\ln k_{A2}$ introduced in Eqs.~\eqref{eq:sigmaA1} and \eqref{eq:sigmaA2} 
can be represented for hydrogenic systems in terms of $\ln k_0({\rm H})$ and $\ln k_3$ \cite{pachucki05} (as shown in Table~\ref{tab:numvalueshe}). 

Comparing Eq.~\eqref{eq:sigmaH} to the result published in Refs.~\cite{yerokhin11,yerokhin12} it differs in the constant term $-421/96$ instead of $-221/64$, due to omitted contributions coming from magnetic moment anomaly. We observe that the new analytic result for hydrogenic systems is in much better agreement with the numerical values from Refs.~\cite{yerokhin11,yerokhin12} which were calculated to all orders in $Z\,\alpha$ but exhibited large numerical uncertainties for $Z<10$. 

{\sl Numerical calculations and results} -- To evaluate the magnetic shielding constant for $^3$He including the QED correction $\sigma^{(5)}$ we represent the wave function and other auxiliary functions used in the calculation of the second-order terms in the basis set of explicitly correlated exponential functions. For example, the ground state wave function is given by
\begin{equation}\label{Basis}
\psi(\vec{r}_1,\vec{r}_2) = (1 + {\cal P}_{12})
\sum_{i=1}^{N} c_i\, e^{-\alpha_i\,r_1 - \beta_i\,r_2 - \gamma_i\,r_{12} } \,,
\end{equation}
where  ${\cal P}_{12}$  exchanges $\vec{r}_1$ with $\vec{r}_2$, and all nonlinear parameters $\alpha_i$, $\beta_i$, and $\gamma_i$ were determined variationally. In order to control the numerical uncertainty, we performed the calculations with several basis sets successively increasing their size by a factor of two; e.g., for the ground-state wave function we used $N = 128$, $256$, and $512$ basis functions. The result obtained in the largest  basis set is accurate to 17 significant digits compared to the benchmark result obtained by Korobov {\em at al.} \cite{Aznabaev:2018}. The method of full optimization used in this work has some advantages in numerical calculations.  
First of all, the numerical uncertainties with these compact wave functions are negligible in comparison 
to uncertainties due to  the omitted higher-order corrections. Moreover, the calculations of all the second-order matrix elements have been performed in quadrupole precision arithmetics, with the full optimization of the intermediate basis of the comparable same size as the ground-state wave function. Similarly, the optimization of the intermediate basis in the calculation of Bethe-type logarithms using integral representations was essential in obtaining high accuracy of numerical results.

\begin{table}[ht]
\caption{Numerical values of He matrix elements. The results are in atomic units and using the notation $\sigma^{(n)}=\alpha^n\tilde{\sigma}^{(n)}$. }\label{tab:numvalueshe}
\begin{center}
\begin{tabular}{ld{3.22}}
    \hline \hline \\
operator &  \centt{value} \\[1ex]
\hline
$E_0$ & -2.903\,724\,377\,034\,119\,59(1)\\
$\left\langle U \right\rangle$ & 3.376\,633\,601\,434\,081(4) \\
$ {\cal D}_{A1} $ & 98.798\,613\,9(3) \\
$ {\cal D}_{A2}$  & 91.002\,103\,1(3)\\[1ex]
$\ln k_0   $      & 4.370\,160\,22(2) \\
$\ln k_0({\rm H})+2\,\ln Z   $  & 4.370\,422\,917   \\[1ex]
$\ln k_{A1}$      & 4.829\,409(3) \\
$\ln k_0({\rm H}) + 2\,\ln Z +1/2 $ & 4.870\,422\,917\\[1ex]
$\ln k_{A2}$      & 4.638\,660(15) \\
$\ln k_3+2\,\ln Z $ & 4.659\,100\,906\\[1ex]
$\tilde\sigma_{A1} $ &  33.750\,67(2) \\
$\tilde\sigma_{A2} $ & -48.007\,69(14) \\
$\tilde\sigma_{B0}$  &  70.054\,125\,1(2) \\
$\tilde\sigma_{B1}$  & -55.342\,119\,09(14) \\
$\tilde\sigma_{B2}$  &   4.188\,033\,454(7) \\
$\tilde\sigma_{B3}$  &   0.011\,67(3)\\
$\tilde\sigma^{(5)} $      &   4.654\,69(15) \\
\hline\hline
\end{tabular}
\end{center}
\end{table}

 From the analysis of convergence, we obtained the extrapolated mean values of the operators. The different contributions to the QED correction of the magnetic shielding in $^3$He are given in Table \ref{tab:numvalueshe}, while the numerical calculation of the lower-order contributions in $\alpha$ is presented in detail in Ref. \cite{rudzinski09}. The numerical value for the shielding Bethe logarithms $\ln k_{A1}$ and $\ln k_{A2}$ were verified in Table \ref{tab:numvalueshe} by the hydrogenic counterparts, 
 because we expect a minor dependence on the number of electrons similarly to  standard Bethe logarithms.
 
 The final results for the shielding in $^3$He are summarized in Table \ref{tab:shieldingstot}. 
 We note that the QED correction is subject to cancellations of the different contributions, and the value presented here for 
$\sigma^{(5)}=96.3\cdot10^{-12}$ is significantly lower than the one previously reported  in Ref. \cite{rudzinski09}, $\sigma^{(5)}_{\mathrm{prev}}=502\cdot10^{-12}$, in which only the leading logarithmic contribution was included. This difference is also reflected in the final value of the shielding. The convergence of the expansions in the fine structure constant $\alpha$ and the electron-nuclear mass ratio $m/m_N$ is very rapid, which justifies our approach based on the NRQED theory. The numerical uncertainties are completely negligible, and only the unknown higher-order terms in $\alpha$ and $m/m_N$ contribute to the uncertainty. The finite nuclear size effects $E_{\rm NS}$ are significant only for heavy elements; here we expect them to be as important as for the binding energy, which for the ground state of He amounts to \cite{pachucki17} $E_{\rm NS}/E_0 \approx 5\cdot 10^{-9}$, which is negligible with respect to the current relative uncertainty of $\sigma$.

\newcommand{\mct}[1]{\multicolumn{2}{l}{#1}}
\begin{table}
\caption{Contributions to the shielding constant $\sigma \cdot 10^6$ for $^1$H, $^3$He$^+$, and $^3$He. 
Quantities that are preceded by ``$\pm$'' represent the uncertainties. Numerical values $\sigma^{(2)}$ and $\sigma^{(4)}$ for $^3$He are taken from Ref.~\cite{rudzinski09}. $\sigma^{(2,2)}$(He) is estimated to be between  $\sigma^{(2,2)}$(He$^+$) and $2\,\sigma^{(2,2)}$(He$^+$).
The relative uncertainty of the finite nuclear mass correction $\sigma^{(4,1)}$ is estimated as $2\,m/m_N$ of $\sigma^{(4)}$.
$\sigma^{(6)}$ is partially known from the Dirac equation \cite{ivanov09}, but we expect cancellation with the radiative correction, 
so we estimate the uncertainty originating from this contribution as $(Z\, \alpha)^2\,\sigma^{(4)}$.}
\label{tab:shieldingstot}
\begin{center}
\begin{tabular}{ld{3.9}d{3.9}d{3.11}}
    \hline \hline & \\
 & \centt{$^1$H} &  \centt{$^3$He$^+$} & \centt{$^3$He} \\[1ex]
\hline
$\sigma^{(2)}$ &17.750\,451\,5 & 35.500\,903\,0 &  59.936\,770\,5  \\
$\sigma^{(4)}$ &0.002\,546\,9 &  0.020\,375\,1 &   0.052\,663\,1  \\
$\sigma^{(5)}$ &0.000\,018\,4 &  0.000\,082\,0 &   0.000\,096\,3\\
$\sigma^{(2,1)}$ & -0.017\,603\,7& -0.013\,933\,4 &  -0.022\,511\,5  \\
$\sigma^{(2,2)}$ &  0.000\,022\,7& 0.000\,007\,1& 0.000\,010\,7(36)   \\
$\sigma^{(4,1)}$ & \pm 0.000\,002\,8 & \pm 0.000\,007\,4& \pm 0.000\,019\,2   \\
$\sigma^{(6)}$ & \pm 0.000\,000\,1 & \pm 0.000\,004\,3  & \pm  0.000\,011\,2  \\[2ex]
$\sigma\cdot10^6$& 17.735\,436(3) & 35.507\,434(9) &   59.967\,029(23)\\[2ex]
\mct{Vaara, Pyykk{\"o} (2003) \cite{vaara03}} & & 59.93(4) \\
\mct{Kudo, Fukui (2005) \cite{Kudo:2005}} & & 59.8 \\
\mct{Antu\v{s}ek {\it et al.} (2007) \cite{Antusek:2007}} & & 59.908\,03 \\
\mct{Rudzi{\'n}ski {\it et al.} (2009) \cite{rudzinski09}} & & 59.967\,43(10) \\
\mct{Seino, Hada (2010) \cite{seino10}} & & 59.95 \\
\mct{Kupka {\it et al.} (2013) \cite{kupka2013}} & &  59.930 \\
\hline\hline& 
\end{tabular}
\end{center}
\end{table}
 
Our result for the magnetic shielding in $^3$He is in disagreement with almost all previous calculations in Refs. \cite{Kudo:2005, Antusek:2007, seino10, kupka2013}, which use the standard quantum chemistry codes and do not present any uncertainties. Moreover, none of them include the finite nuclear mass or the QED effects, but the differences are significantly greater than the omitted corrections. The only exception is the results of Vaara and Pyykk{\"o} \cite{vaara03}, which, although the oldest one, present uncertainties in agreement with  our result. 

{\sl Conclusions} -- We have presented the complete formula of the leading quantum electrodynamics correction to the magnetic shielding for hydrogen- and helium-like atomic systems, and performed calculations for $^1$H and the particularly important cases of $^3$He$^+$ and $^3$He, which are accurate to $3\cdot10^{-12}$, $9\cdot10^{-12}$, and $23\cdot10^{-12}$, respectively. The results permit establishing a new standard for the absolute magnetometry based on $^3$He NMR probes, by removing one of the bottlenecks in determining $\mu_h$ by indirect methods. 

Moreover, the NMR frequency ratio of $^3$He and of H$_2 $ is known with a relative accuracy of about $10^{-9}$ \cite{garbacz2012} and $\mu_p$ to $2.9\cdot 10^{-10}$, so it is enough to provide the magnetic shielding of H$_2$  with about $10^{-5}$ relative accuracy, to obtain
$\mu_h$ with about $10^{-9}$ accuracy for verifying the consistency among various determinations. This would require the calculation of only nonadiabatic and relativistic effects, which we plan to perform this in the near future, while the QED corrections, estimated using the atomic value in Table \ref{tab:shieldingstot}, are negligible.

The result of the magnetic shielding in $^3$He$^+$ can be used for the determination of $\mu_h$ from the measurement of $^3$He$^+$  \cite{mooser18,schneider19}, being close to completion \cite{private}. Alternatively, if $\mu_h$ is measured directly by comparison with the cyclotron frequency \cite{schneider19}, the result for the $^3$He magnetic shielding will provide an accurate magnetic moment of atomic $^3$He for use in absolute magnetometry. Finally, the NRQED approach, presented in this work, to calculate the magnetic shielding, 
can be extended to nuclei of other light atomic systems, such as Li, Be, and B, which will lead to the determination of their magnetic moments with the accuracy of experiments, which for the case of Be$^+$ \cite{wineland1983} is about $10^{-9}$, i.e. higher 
by a few orders of magnitude in comparison to the presently known value of $\mu_{\rm Be}$ \cite{antusek2013,pachucki2010}.

\begin{acknowledgments}
D.W. thanks F. Merkt for his unconditional support to work on this project. This research was supported by National Science Center (Poland) Grants No. 2017/27/B/ST2/02459 and 2019/35/N/ST4/04445, as well as by a computing grant from the Pozna\'n Supercomputing and Networking Center and by PL-Grid Infrastructure.
\end{acknowledgments}



%

\end{document}